\documentclass[twocolumn,preprintnumbers,amsmath,amssymb]{revtex4}
\input psfig.sty  
\def \be {\begin{equation}} 

\def \ee {\end{equation}} 
\def \bea {\begin{eqnarray}} 
\def \eea {\end{eqnarray}} 

\usepackage{graphicx}
\usepackage{dcolumn}
\usepackage{bm}
\usepackage{epsfig}

\newcommand*{\ltsim}{\ {\raise-.75ex\hbox{$\buildrel<\over\sim$}}\ }
\newcommand*{\gtsim}{\ {\raise-.75ex\hbox{$\buildrel>\over\sim$}}\ }
\newcommand*{\proptosim}{\ {\raise-.75ex\hbox{$\buildrel\propto\over\sim$}}\ }

\begin{document}

\title{Dynamical and observational analysis of interacting models}

\author{F. E. M. Costa$^{1}$} \email{ernandesmatos@ufersa.edu.br}
\author{A. O. Ribeiro$^{2}$} \email{anderson@on.br}
\author{F. Roig$^{3}$} \email{froig@on.br}

\affiliation{$^1$Universidade Federal Rural do Semi-\'Arido, 59900-000, Pau dos Ferros, RN, Brazil}
\affiliation{$^2$Centro Universit\'ario Geraldo Di Biase, 27101-970, Barra do Pira\'{\i}, RJ, Brazil}
\affiliation{$^3$Observat\'orio Nacional, 20921-400, Rio de Janeiro, RJ, Brazil}

\date{\today}

\begin{abstract}

We investigate the dynamical behaviour of a general class of interacting models in the dark sector in which the 
phenomenological coupling between cold dark matter and dark energy is a power law of the cosmic scale factor. 
From numerical simulations we show that, in this background, dark energy always dominates the current composition 
cosmic. This behaviour may alleviate substantially the coincidence problem. By using current type Ia supernovae, 
baryonic acoustic oscillations and cosmic microwave background data, we perform a joint statistical analysis and 
obtain constraints on free parameters of this class of model.\\

\end{abstract}

\maketitle

\section{Introduction}

A considerable number of observational data such as cosmic microwave background (CMB) (Spergel et al. 2007), large scale structure surveys (Eisenstein et al. 2005) and type Ia supernovae (SN Ia) (Permulter et al. 1998; Riess et al. 1998), when combined, 
indicate that Universe has a spatially flat geometry and is undergoing an accelerated expansion phase. 
Considering that the general relativity describes the gravity on large scales and that the space-time is 
homogeneous and isotropic, then we must assume the existence of a new hypothetical energy component with 
negative pressure, the so-called dark energy, that dominates the current composition of the cosmos 
[see, e.g., (Sahni \& Starobinsky 2000; Padmanabhan 2003) for some recent reviews on this topic].

The vacuum state of all existing fields in the Universe, that acts in the Einstein field equations as a 
cosmological constant $\Lambda$, is the simplest and most natural candidate to dark energy. Flat models with 
a very small cosmological term provide a very good description of the observed Universe. However, the $\Lambda$ 
value inferred by observations ($\rho_{\Lambda} = \Lambda/8\pi G \lesssim 10^{-47}$ ${\rm{GeV}}^4$) differs from 
theoretical estimates given by quantum field theory ($\rho_{\Lambda}  \sim 10^{71}$ ${\rm{GeV}}^4$) by almost 
120 orders of magnitude. This large discrepancy originates an extreme fine-tuning problem and requires a 
complete cancellation from an unknown physical mechanism. The difficulty in explaining this cancellation is 
known as the cosmological constant problem (Weinberg 1989).

Another problem with $\Lambda$ (and which also persists in dark energy models) is to understand why dark energy 
density is not only small, but also of the same
order of magnitude of the energy density of cold dark matter. Since both components 
(dark energy and dark matter) are usually assumed to
be independent and, therefore, scale in different ways, this would require an unbelievable coincidence, 
the so-called coincidence problem (CP).

From the theoretical viewpoint, the CP could be solved if we knew some physical mechanism that leads the relative 
densities (in units of the critical density) of dark
matter $(\Omega_{dm})$ and dark energy $(\Omega_{x})$ to similar values at the current time.
From the phenomenological viewpoint, the coincidence problem is alleviated allowing that dark matter and 
dark energy to interact. This phenomenology in turn gave origin
to the so-called models of coupled quintessence, which have been largely explored in the literature (Amendola 2000; Chimento et al. 2003; Costa \& Alcaniz 2010; Costa 2010; Costa, Alcaniz \& Deepak 2012; Costa 2017). These scenarios 
are based on the premise that, unless some special and unknown symmetry in nature prevents or 
suppresses a non-minimal coupling between dark matter and dark energy,
a small interaction cannot be ruled out [see (Carroll 1998) for a discussion].

In particular, two conditions must be met to solve the coincidence problem: (i) the ratio 
$\Omega_{x}/ \Omega_{dm} = O(1)$ and (ii) the second derivative of
the scale factor must be positive $\ddot{a} > 0$ (Caldera et al. 2008). In other words, this amounts to saying that the coupling between dark matter and dark energy should lead to an accelerated scaling attractor solution.

In this paper we explore the dynamic behaviour of a general class of coupled quintessence models in which the coupling 
between in the dark sector is a power law of the scale factor. By using numerical simulations we show that this class 
of interacting models is not sensitive to the initial conditions and always leads the Universe to a current accelerated 
phase. We also test the observational viability in light of recent type Ia supernovae (SNe Ia) measurements, as given 
by Union 2.1 of the Supernova Cosmology Project (SCP) (Suzuki et al. 2012),
baryon acoustic oscillations (BAO) at three different redshifts $z = 0.20$, and $z = 0.35$ and $z = 0.6$ (Blake et al. 2011) and the shift parameter from the three-year Wilkinson Microwave Anisotropy Probe (WMAP) data (Komatsu et al. 2009).

\section{Dynamics analysis}

Let us consider that the main contributions to the total energy-momentum tensor of the cosmic fluid are non-relativistic 
matter (baryonic plus dark) and a negative-pressure dark energy component. Thus
\begin{equation}
{T}^{\mu \nu} = {T}_{b}^{\mu \nu} + {T}_{dm}^{\mu \nu} + {T}_{x}^{\mu \nu},
\end{equation}
where ${T}_{b}^{\mu \nu}$, ${T}_{dm}^{\mu \nu}$ and ${T}_{x}^{\mu \nu}$ are, respectively, the energy-momentum tensors 
baryonic matter, dark matter and dark energy. By assuming the Friedmann-Lemaitre-Robertson-Walker space-time and a 
coupling in the dark sector, the condition $\nabla_{\nu} {T}^{\mu \nu} = 0$, implies that
\begin{equation}\label{coup0}
\dot{\rho}_{dm} + 3 \frac{\dot{a}}{a}\rho_{dm} = -\dot{\rho}_x -
3\frac{\dot{a}}{a}(\rho_x + p_x )\;,
\end{equation}
and 
\begin{equation}
\dot{\rho}_{b} + 3\frac{\dot{a}}{a}{\rho}_{b} = 0\;.
\end{equation}
where $\rho_{dm}$, $\rho_x$ and $\rho_b$ are the energy densities of the dark matter, dark energy and baryonic 
matter, respectively, whereas $p_{x}$ is the dark energy pressure.
Now, by considering that the dark energy satisfies an equation of state $p_x=\omega \rho_x$, with 
$\omega = \rm{constant} < 0$ and making $N = \ln a$, above equations can be rewritten as
\begin{equation}\label{coup1}
\frac{d{\rho}_{x}}{dN} + 3 (1+ \omega){\rho}_x = -Q\;,
\end{equation}
\begin{equation}\label{coup2}
\frac{d{\rho}_{dm}}{dN} + 3{\rho}_{dm} = Q\;,
\end{equation}
\begin{equation}\label{coup3}
\frac{d{\rho}_{b}}{dN} + 3{\rho}_{b} = 0\;,
\end{equation}
where $Q$ is the coupling function in the dark sector.

By introducing the following variables:
\begin{equation}\label{def5}
X \equiv \frac{8\pi G{\rho}_{x}}{3H^{2}}, \quad \quad Y \equiv \frac{8\pi G{\rho}_{dm}}{3H^{2}}, \quad \quad Z^{2} \equiv \frac{8\pi G{\rho}_{b}}{3H^{2}},
\end{equation}
where $H \equiv \dot{a}/a$ is the Hubble parameter. Note that, we define $X$ and $Y$ and not, say, $X^2$ and $Y^2$ as 
variables to naturally allow for negative $\rho_x$ and $\rho_{dm}$ which leads to more complete understanding
of the dynamics involved. In terms of these new variables Eqs. (\ref{coup1}), (\ref{coup2}) and (\ref{coup3}) can 
be rewritten as
\begin{equation}
\frac{d}{dN}X = 3\omega X(X-1) - \tilde{Q}\;,
\end{equation}
\begin{equation}
\frac{d}{dN}Y = 3\omega XY + \tilde{Q}\;,
\end{equation}
\begin{equation}
\frac{d}{dN} Z = \frac{3}{2}\omega XZ\;,
\end{equation}
where $\tilde{Q} = Q/3H^{2}$.

Now, as we do not know the nature of dark components, it is not possible to derive from first principles the functional
form for the coupling function. Thus, we must assume an appropriated relation for $Q(a)$ or equivalently $\tilde{Q}(N)$. 
Certainly, among many possible functional forms, a very simple choice is
\begin{equation}
Q(a)=\epsilon_{0}a^{\xi}\rho_{dm}\Longrightarrow\tilde{Q}(N)=\epsilon_{0}\left(1+\xi N\right)\mathbf{e}^{\xi N}Y\;.\label{eq:11}
\end{equation}
Therefore, the evolution of this interacting model is described by the following non autonomous system
\begin{equation}
\frac{dX}{dN}=3\omega X\left(X-1\right)-\epsilon_{0}\left(1+\xi N\right)\mathbf{e}^{\xi N}Y\label{inicial}\;,
\end{equation}
\begin{equation}
\frac{dY}{dN}=3\omega XY+\epsilon_{0}\left(1+\xi N\right)\mathbf{e}^{\xi N}Y\label{dmatter}\;,
\end{equation}
\begin{equation}
\frac{dZ}{dN}=\frac{3}{2}\omega XZ\label{baryons}\;,
\end{equation}
that fulfils the condition $X+Y+Z^{2}=1$.

\begin{figure*}
\centerline{\psfig{figure=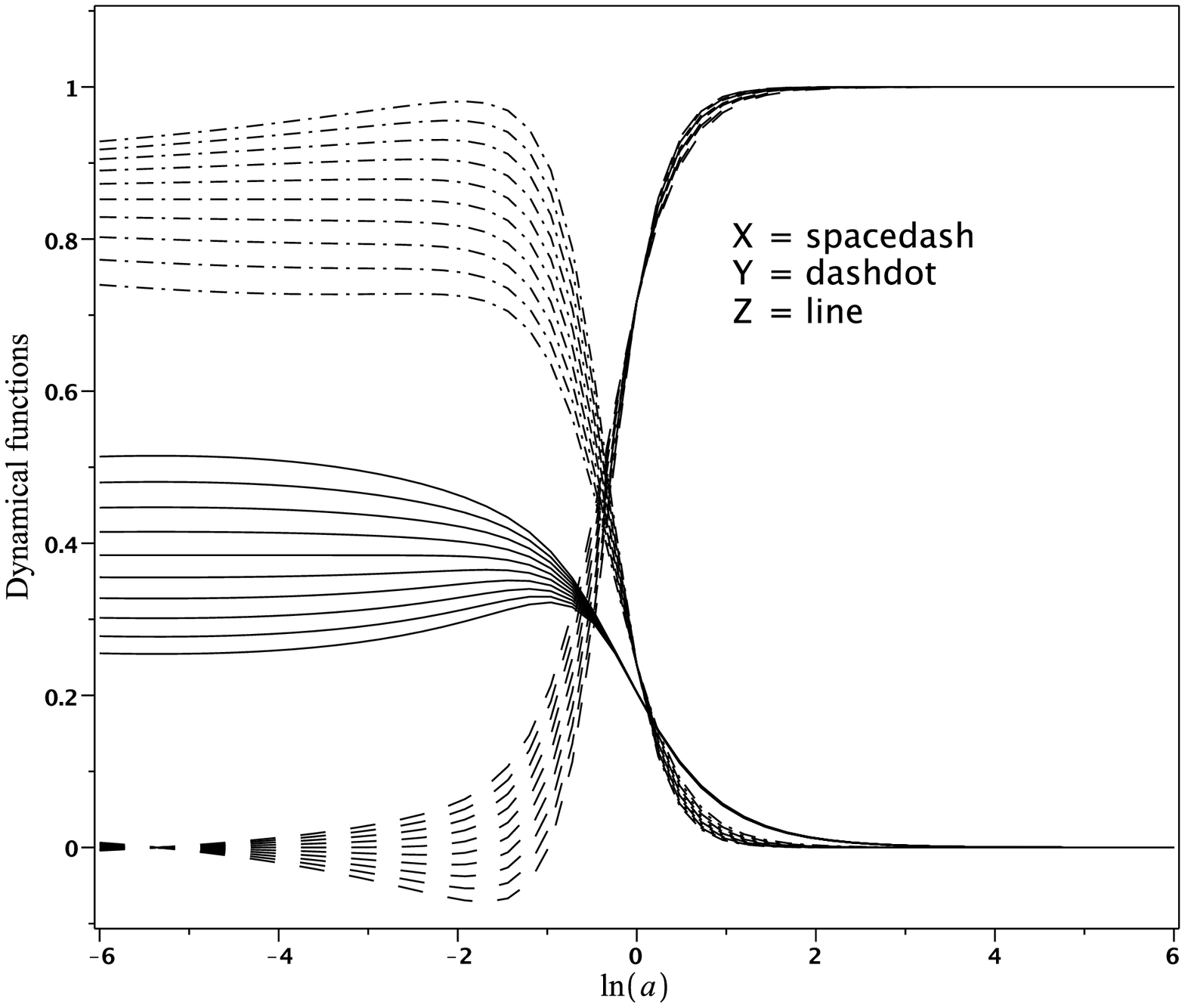,width=2.4truein,height=2.5truein,angle=0}
\psfig{figure=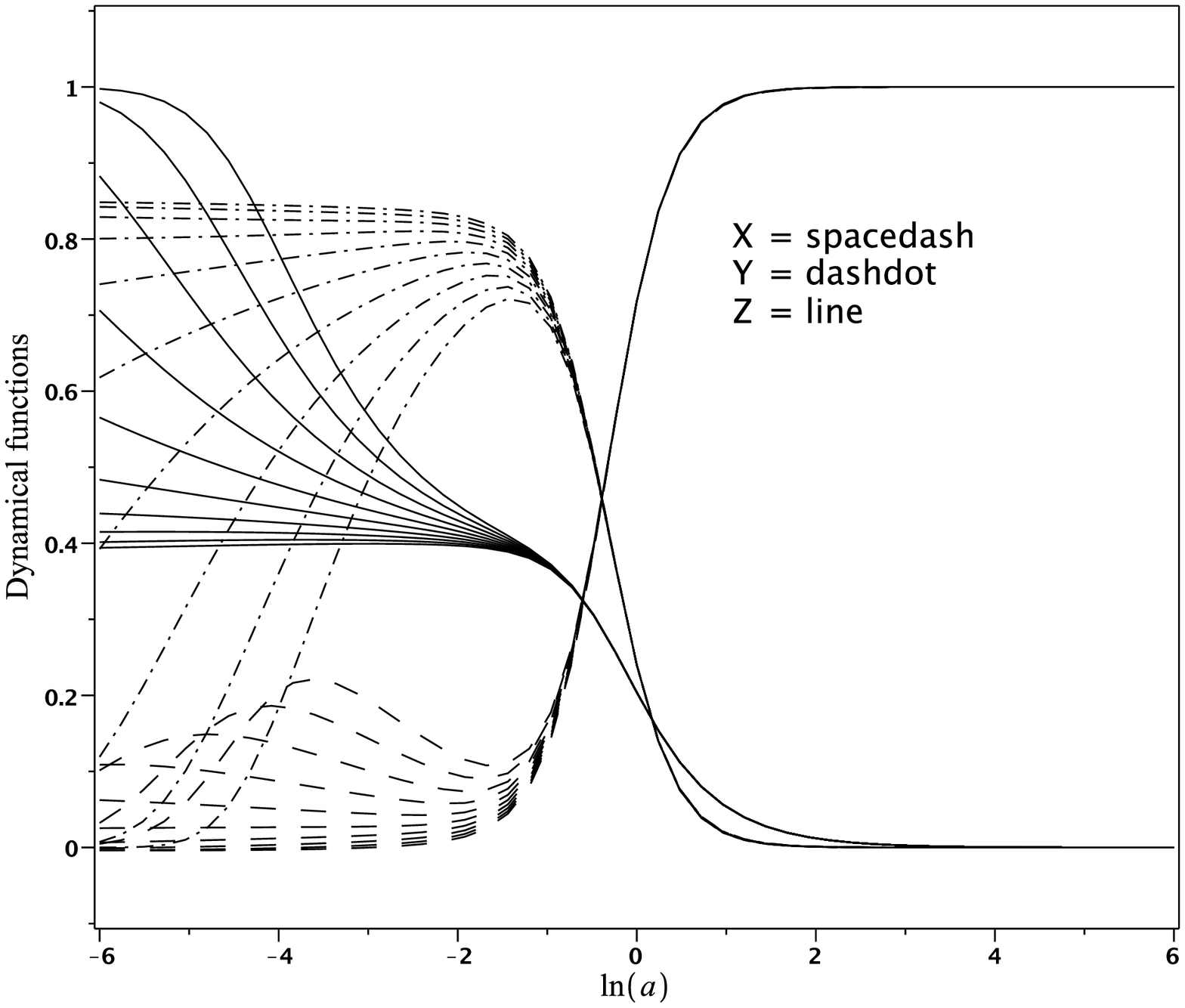,width=2.4truein,height=2.5truein,angle=0}
\psfig{figure=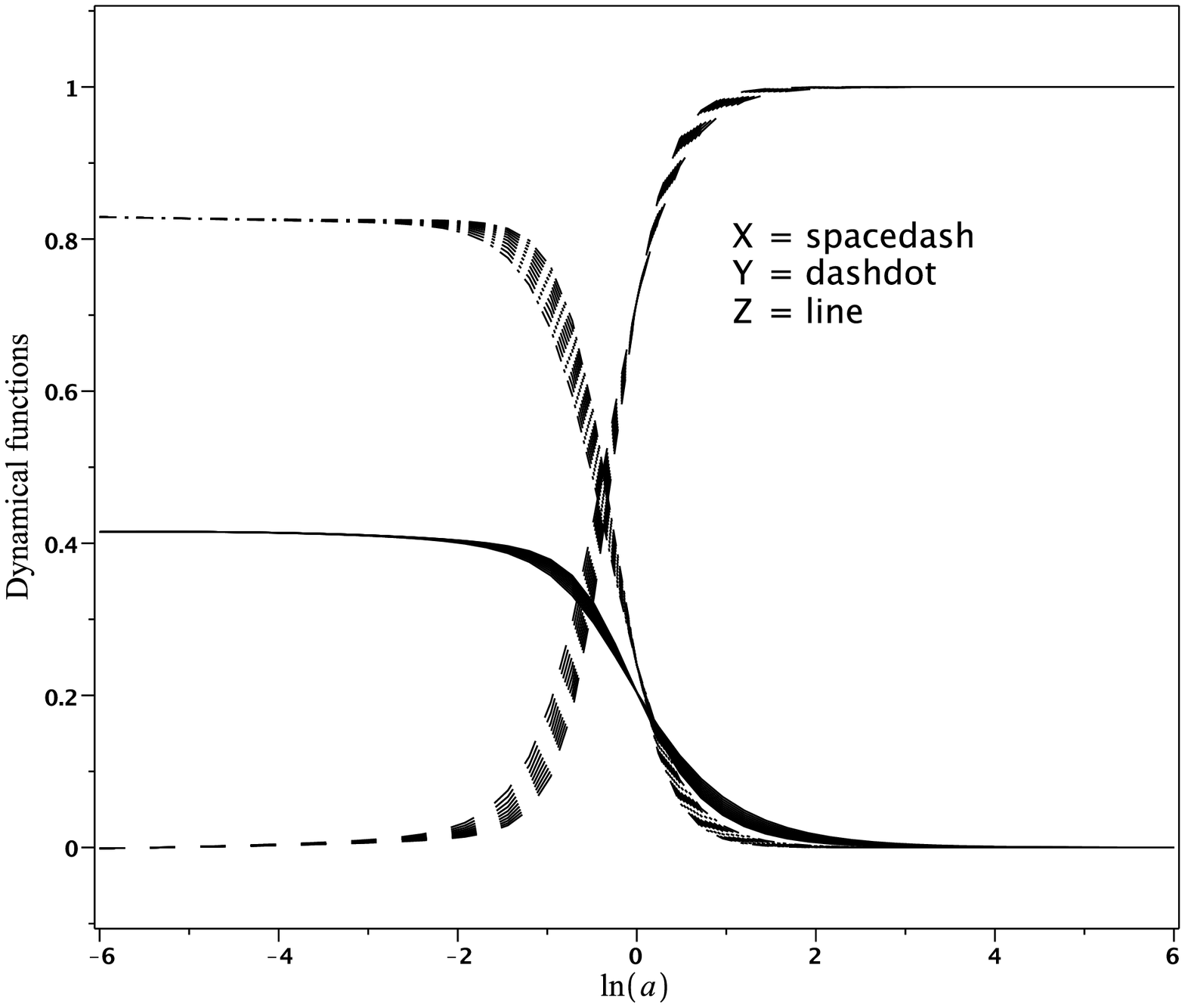,width=2.4truein,height=2.5truein,angle=0}}
\caption{Evolution of the $X,Y,Z$ components of the dynamical system Eqs. (\ref{inicial})-(\ref{baryons}), as a function of $N=\ln(a)$, for $\xi = 0.2$, $\omega = -1$ and $-0.5 \leq \epsilon_0 \leq 0.5$ (left);
$\epsilon_0 = 0.1$, $\omega = -1$ and $-0.5 \leq \xi \leq 0.5$ (middle); and $\epsilon_0 = 0.1$, $\xi = 0.2$ and $-1.2 \leq \omega \leq -0.8$ (right).}
\label{fig1}
\end{figure*}

\subsection{Critical points}

We consider now the critical conditions of Eqs. (\ref{inicial})-(\ref{baryons})
for which ${dX}/{dN}={dY}/{dN}={dZ}/{dN}=0$ for every $N$. There are four of such conditions:
\begin{eqnarray}
C_{1} & : & (X_{1},Y_{1},Z_{1})=(1,0,0)\;,\nonumber \\
C_{2} & : & (X_{2},Y_{2},Z_{2})=(0,0,1)\label{eq:16}\;,\\
C_{3} & : & (X_{3},Y_{3},Z_{3})=(0,0,-1)\;,\nonumber \\
C_{4} & : & (X_{4},Y_{4},Z_{4})=(X^{*},1-X^{*},0)\;,\nonumber 
\end{eqnarray}
where ${\displaystyle X^{*}=-\epsilon_{0}\left(1+\xi N\right)\mathbf{e}^{\xi N}}/{3\omega}$.
The third condition is not of cosmological interest since a negative
density of baryonic matter is physically meaningless, so it will not be considered.
Conditions $C_{1}$ and $C_{2}$ represent fixed points of the system. $C_{1}$ corresponds to 
a dark energy dominated epoch (de Sitter point) while $C_{2}$ corresponds to an epoch dominated by baryonic 
matter only. For $\xi\neq0$, condition $C_{4}$ is neither a fixed point nor a
stationary solution of the system, but it represents a critical point
that moves in the $X,Y$ plane along the line $Y=1-X$. 
The stability of this critical point will then affect the behaviour of the solutions
of the system in the neighbourhood of this line. This critical point is of cosmological interest 
because it provides the transition between the past dark matter dominated epoch and the present 
cosmic acceleration. We may note that,
since $\omega<0$ always, then for $\epsilon_{0}>0$ there will be
some $N=N_{1}$ for which $X^{*}=1$. In other words, the Universe will always pass through a dark energy dominated epoch. 
For values of $N$ around $N_{1}$, the stability of the critical point $C_{4}$ must be compatible
with the stability of the fixed point $C_{1}$.

To analyse the stability of these critical points, we perform a linearisation
of the system around each point $C_{i}$ to get the variational equations:
\begin{equation}\label{eq:17}
\frac{d\mathbf{E}}{dN}=\mathbf{M}(N)\,\mathbf{E}\;,
\end{equation}
where $\mathbf{E}$ is a column vector with components $X-X_{i}$,
$Y-Y_{i}$, $Z-Z_{i}$, and $\mathbf{M}(N)$ is the Jacobian matrix
of the system evaluated at the critical point $C_{i}$ 
\[
\mathbf{M}(N)=\left(\begin{array}{ccc}
3\omega(2X_{i}-1) & -f(N) & 0\\
3\omega Y_{i} & 3\omega X_{i}+f(N) & 0\\
{\displaystyle \frac{3}{2}\omega Z_{i}} & 0 & {\displaystyle \frac{3}{2}\omega X_{i}}
\end{array}\right)
\]
with $f(N)=\epsilon_{0}\left(1+\xi N\right)\mathbf{e}^{\xi N}$. Since
Eq. (\ref{eq:17}) is a linear non autonomous system, the classical
eigenvalues analysis that is valid for autonomous systems cannot be
applied here, and we have to compute the characteristic exponents
of this system form the definition:
\[
\lambda=\lim_{N\rightarrow\infty}\frac{1}{N}\ln\left|\mathbf{E}(N)\right|.
\]
It is worth noting that the stability of the fixed points may (and will) vary with $N$
and that the above definition allows to characterize only the 
asymptotic stability. In the following, it is assumed that $\epsilon_{0},\xi$ and $\omega$
are all bounded values.

Around the fixed point $C_{1}$, the general solution $\mathbf{E}(N)$
of Eq. (\ref{eq:17}) is given by:
\begin{eqnarray*}
E_{X}(N) & = & E_{X0}\mathbf{e}^{3\omega N}+E_{Y0}\mathbf{e}^{3\omega N}-E_{Y0}\mathbf{e}^{3\omega N+g(N)}\;,\\
E_{Y}(N) & = & E_{Y0}\mathbf{e}^{3\omega N+g(N)}\;,\\
E_{Z}(N) & = & E_{Z0}\mathbf{e}^{\frac{3}{2}\omega N}\;,
\end{eqnarray*}
with ${\displaystyle g(N)=\int f(N)dN=\epsilon_{0}N\mathbf{e}^{\xi N}}$.
If $\xi<0$, the characteristic exponents are: $\lambda_{x,y} = 3\omega $ and $\lambda_{z} = 3\omega /2$
which are all negative, so the fixed point is asymptotically stable,
more precisely an attractor.
On the other hand, if $\xi>0$, then at least one characteristic exponent
diverges and the fixed point is stable (attractor) if $\epsilon_{0}\leq0$ and unstable (saddle) otherwise. 
Since $\xi\neq0$ appears always multiplying $N$, then a reversion in the sign of $N$ implies a reversion of
the stability of the fixed point, i.e. if the fixed point
is an attractor for past times, it is a repulsor for future times
and vice-versa. For $\xi=0$, the first
two characteristic exponents are $\lambda_{x,y}=3\omega+\epsilon_{0}$
and the point will be an attractor if $\epsilon_{0}<-3\omega$ and a saddle if $\epsilon_{0}>-3\omega$. 

\begin{figure*}
\centerline{%
\psfig{figure=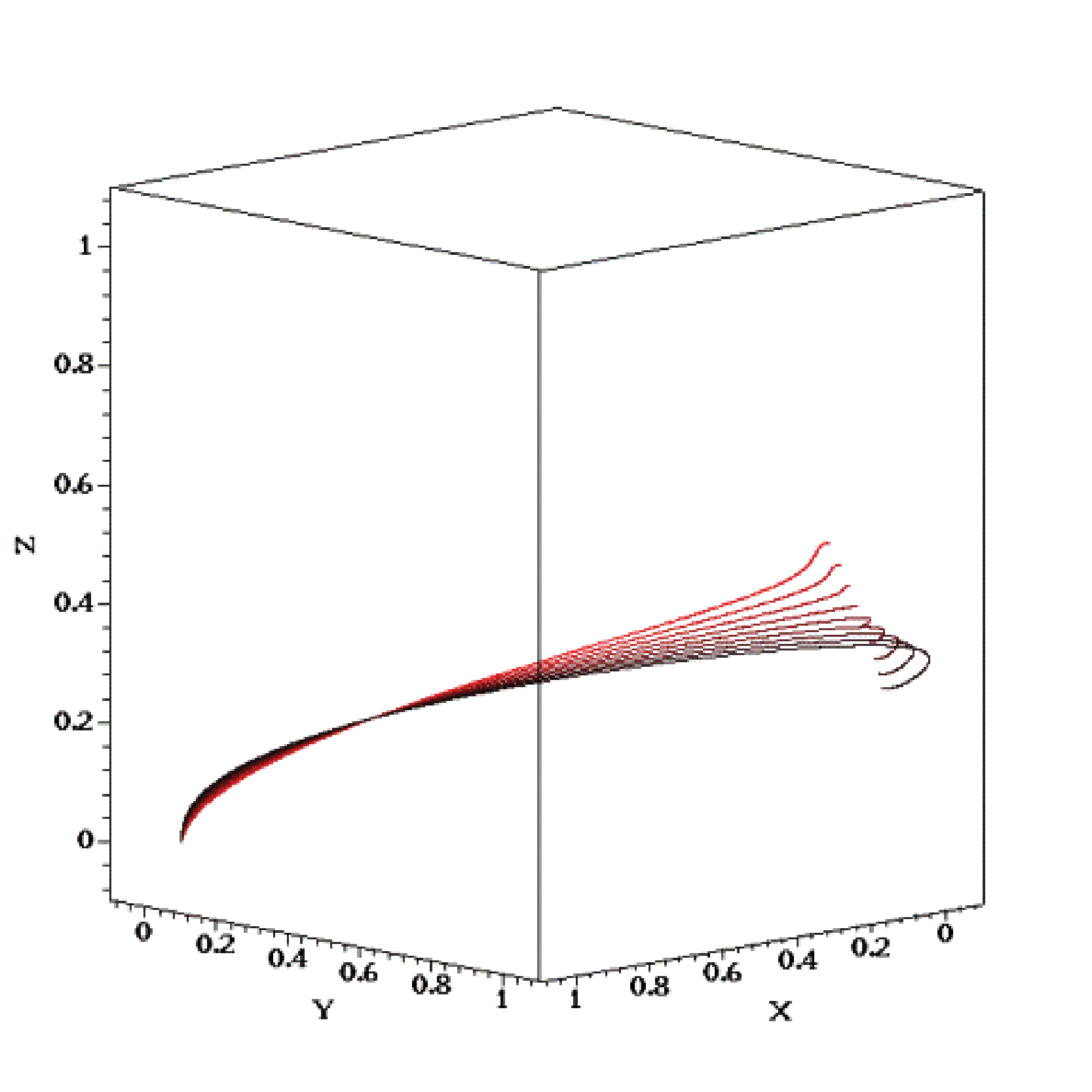,width=0.33\textwidth,angle=0}
\psfig{figure=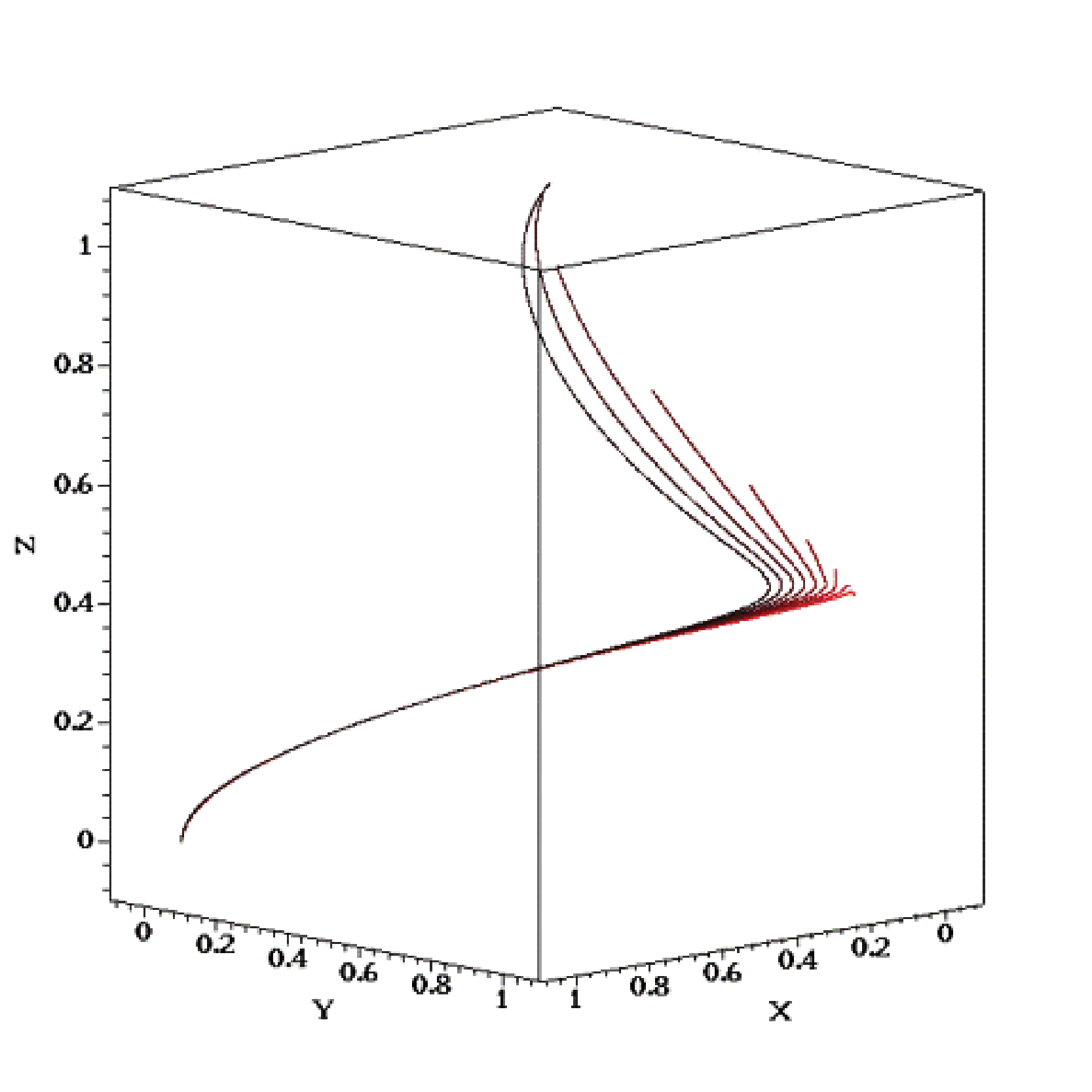,width=0.33\textwidth,angle=0}
\psfig{figure=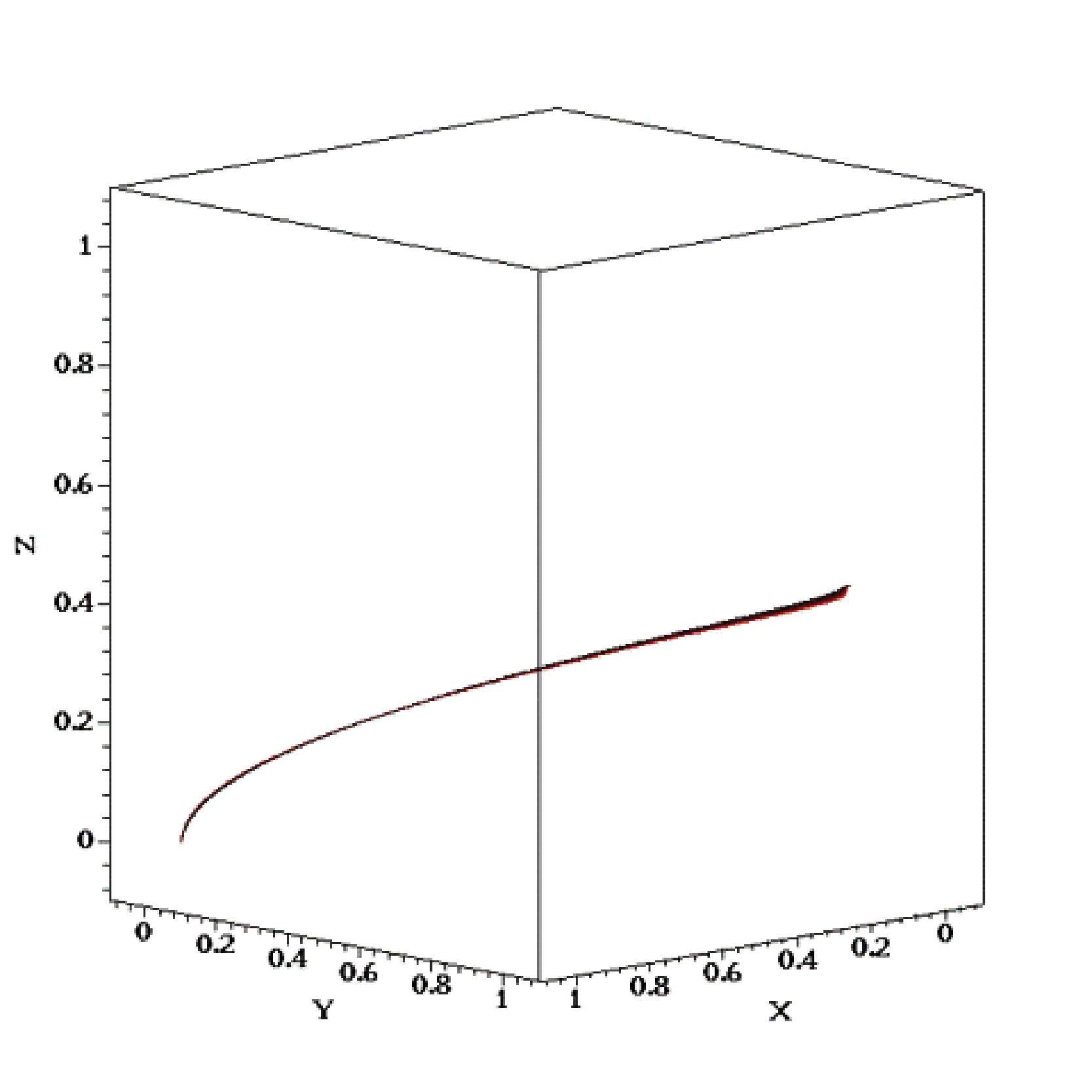,width=0.33\textwidth,angle=0}} 
\vspace{0.5cm}
\centerline{%
\psfig{figure=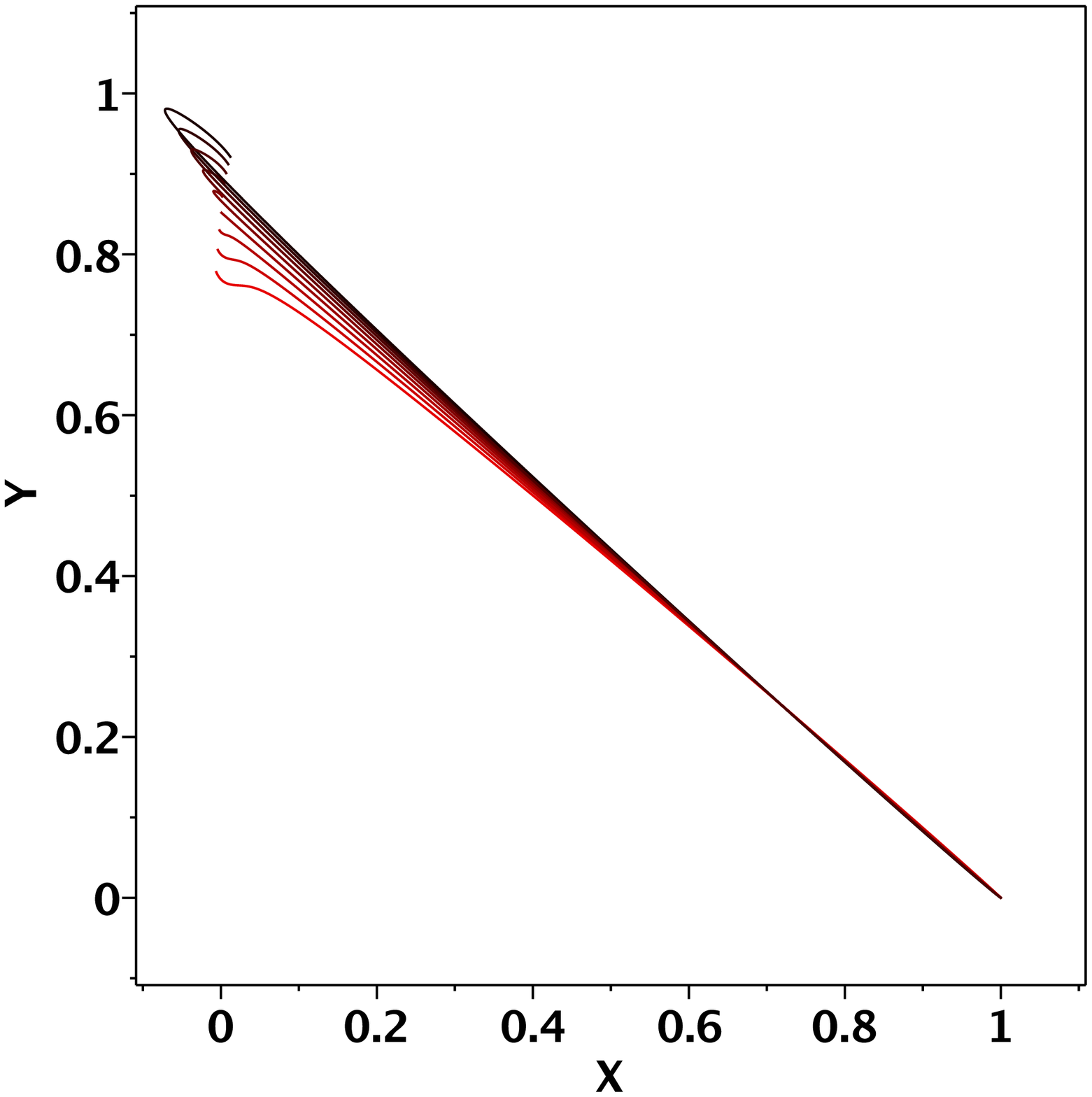,width=0.33\textwidth,angle=0}
\psfig{figure=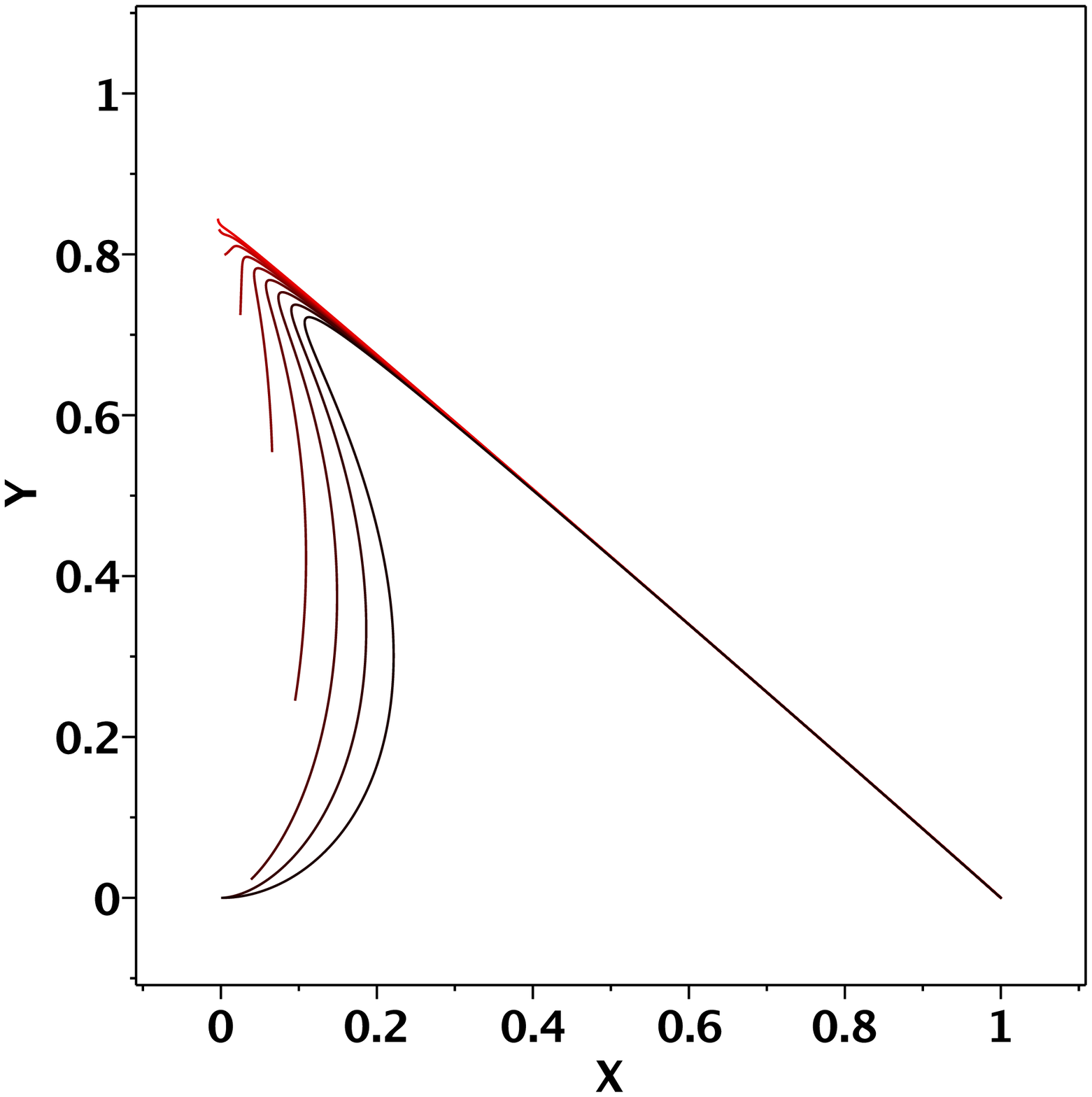,width=0.33\textwidth,angle=0}
\psfig{figure=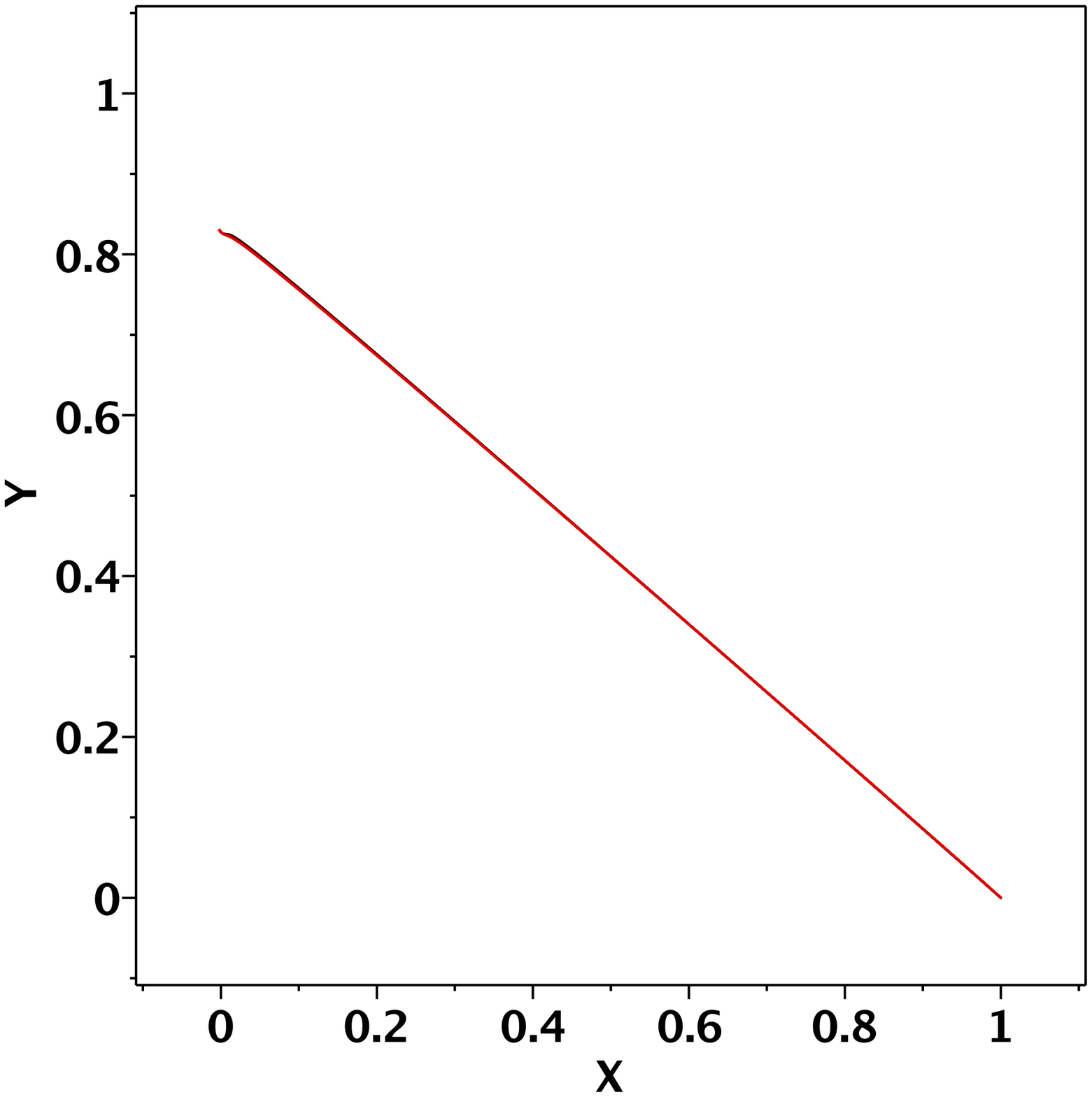,width=0.33\textwidth,angle=0}}
\caption{The 3D phase space ($X,Y,Z$) (above) and 2D projection ($X,Y$) (below) of the solutions shown in Fig. \ref{fig1}.}
\label{fig2}
\end{figure*}

Around the fixed point $C_{2}$, it is enough to analyse the solution
of Eq. (\ref{eq:17}) for $E_{X},E_{Y}$ to conclude that this point is 
asymptotically unstable for all $N$:
\begin{eqnarray*}
E_{X}(N) & = & E_{X0}\mathbf{e}^{-3\omega N}-E_{Y0}\mathbf{e}^{-3\omega N}p(N)\;,\\
E_{Y}(N) & = & E_{Y0}\mathbf{e}^{g(N)}\;,
\end{eqnarray*}
with ${\displaystyle p(N)=\epsilon_{0}\int\mathbf{e}^{3\omega N+g(N)+\xi N}\left(1+\xi N\right)dN}$.
We note that if $\xi>0$, then $\lambda_{y}$ is divergent and the
fixed point is unstable. If $\xi\leq0$, the integral $p(N)$ is convergent for 
$N\rightarrow\infty$ and $\lambda_{x}=-3\omega>0$,
thus the fixed point is still unstable, actually a global repulsor.  

Finally, around the critical point $C_{4}$, the general solution
of Eq. (\ref{eq:17}) is
\begin{eqnarray*}
E_{X}(N) & = & E_{X0}\mathbf{e}^{-3\omega N-g(N)}-(E_{X0}+E_{Y0})\mathbf{e}^{\xi N-g(N)}q(N)\;,\\
E_{Y}(N) & = & (E_{X0}+E_{Y0})\mathbf{e}^{-g(N)}-E_{X}(N)\;,\\
E_{Z}(N) & = & E_{Z0}\mathbf{e}^{-g(N)/2}\;,
\end{eqnarray*}
with ${\displaystyle q(N)=\epsilon_{0}\left[\frac{1}{3\omega+\xi}+\frac{\xi^{2}N+3\xi\omega N-\xi}{\xi^{2}+6\xi\omega+9\omega^{2}}\right]}$. Again, if $\xi>0$ at least one characteristic exponent is divergent and if $\xi<0$ at least one characteristic exponent is positive,
therefore the critical point is always asymptotically unstable. In particular, if
$\epsilon_{0}>0$ then the critical point is a saddle in the $X,Y$
plane that travels from the point (0,1) to (1,0) if $\xi>0$, or from
(1,0) to (0,1) if $\xi<0$, which is compatible with 
the stability of the $C_{1}$ point. If $\xi=0$, the critical point $C_{4}$
becomes a fixed point of Eqs. (\ref{inicial})-(\ref{baryons}), and it
is saddle if $\epsilon_{0}<-3\omega$ and an attractor $\epsilon_{0}>-3\omega$.

From the above analysis, we conclude that the stability of the critical
points is strongly dependent on the sign of the parameter $\xi$ and
on the interval of the independent variable $N$ under consideration.
As we will see in the following, for the typical range of variation
of the model parameters and of the independent variable, the fixed
point $C_{1}$ is always an attractor

\subsection{Dynamical evolution}

We have simulated the evolution of the dynamical system for different values
of the model parameters. The solutions have been obtained by numerical
integration of Eqs. (\ref{inicial})-(\ref{baryons}) using an adaptive-step 4th. 
order Runge-Kutta algorithm. All the simulations started
from initial conditions at $N=0$ compatible with the present values
of the dark and baryonic density components of the system ($X=0.7184$,
$Y=0.24$, $Z=0.204$). The simulations spanned the interval $-6.0\leq N\leq 6.0$,
corresponding to redshifts $100\leq z\leq -0.99$.
The parameters of the model $\epsilon_{0},\omega,\xi$ were varied
within the following intervals: $-0.5\leq\epsilon_{0}\leq0.5$, $-1.2\leq\omega\leq-0.8$,
$-0.5\leq\xi\leq0.5$. 

\begin{figure*}
\centerline{\psfig{figure=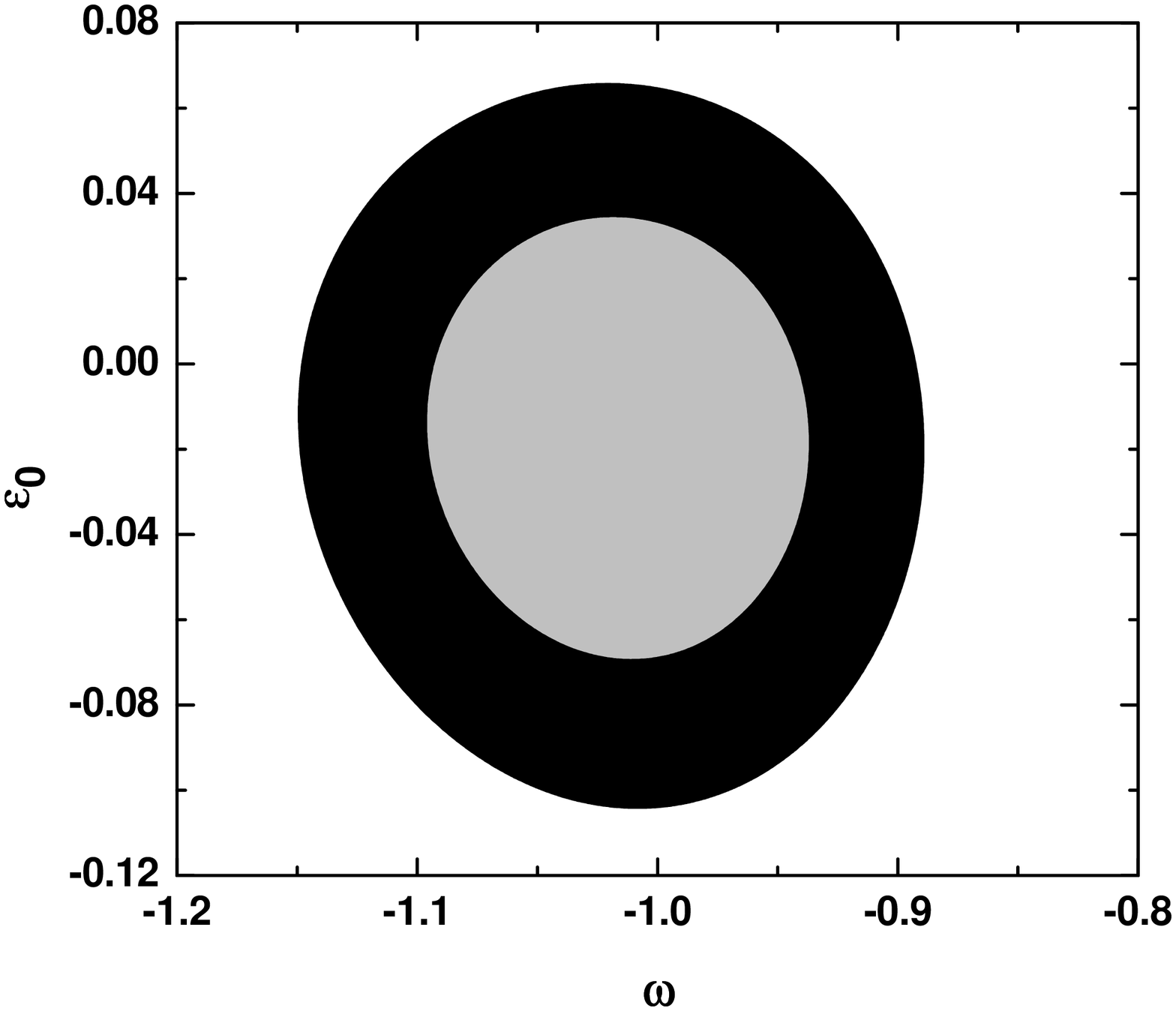,width=2.4truein,height=2.5truein,angle=-90}
\psfig{figure=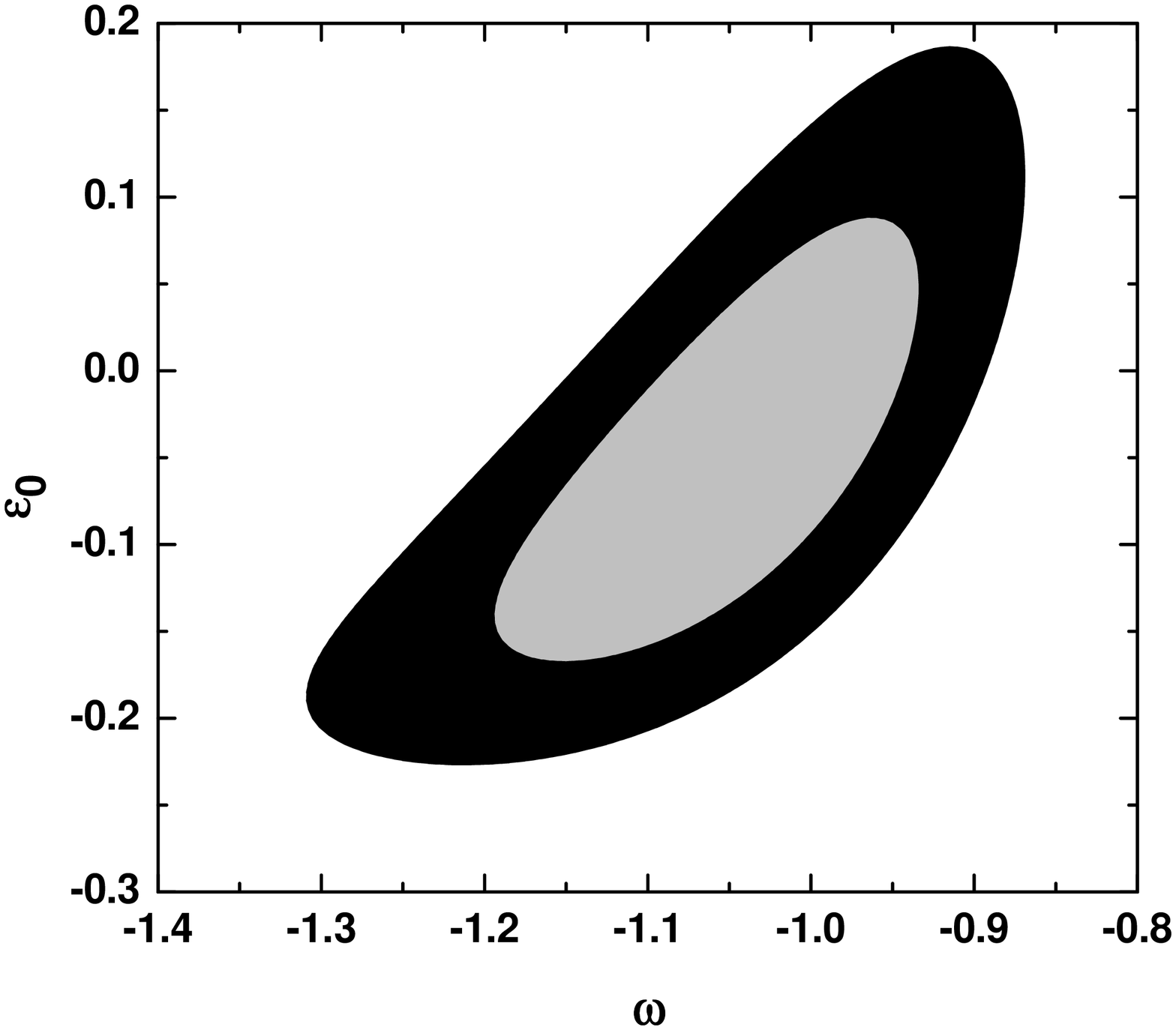,width=2.4truein,height=2.5truein,angle=-90}
\psfig{figure=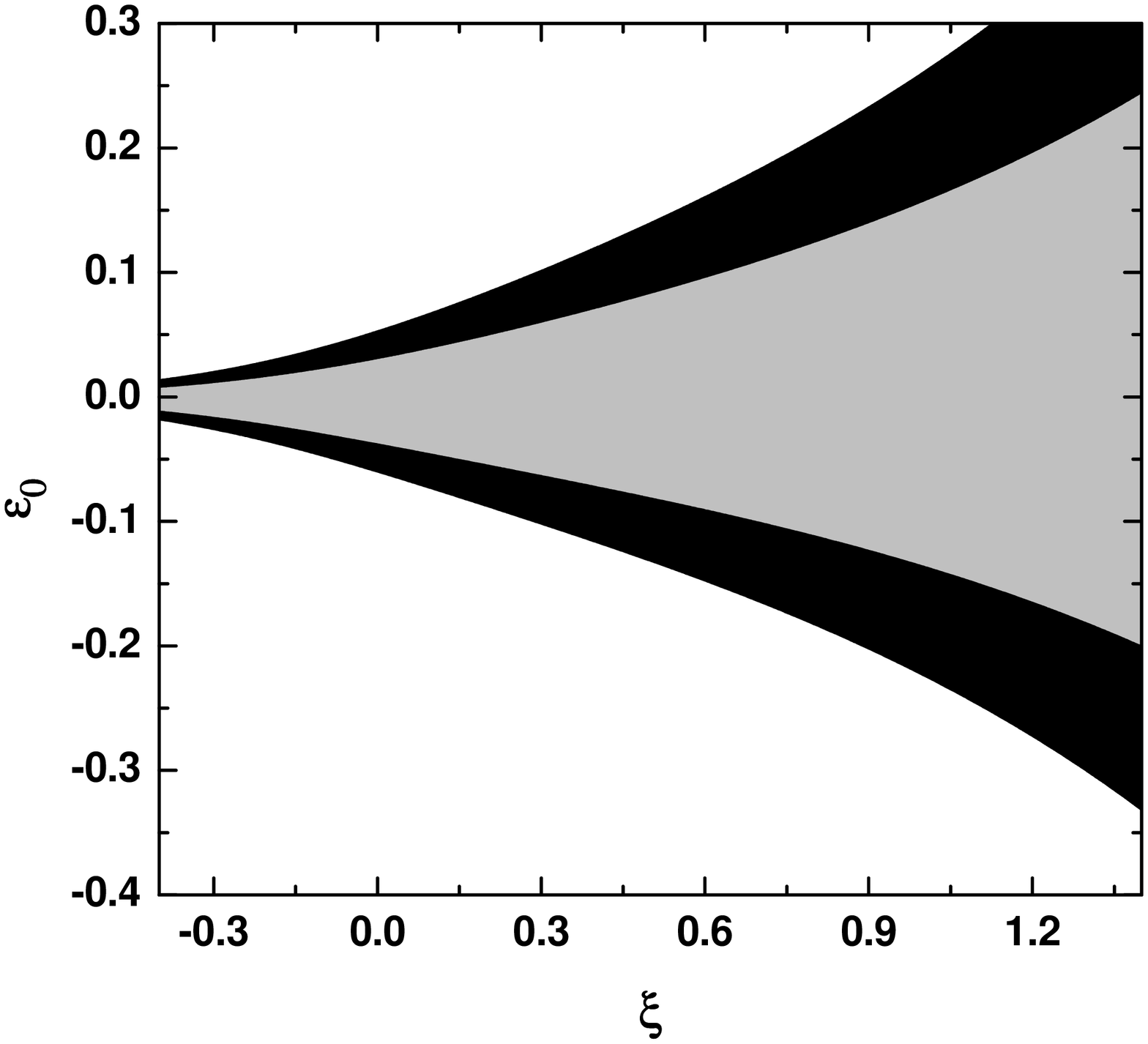,width=2.4truein,height=2.5truein,angle=-90}}
\caption{Results of our statistical analysis. Contours of $\chi^2$ in the planes $\omega - \epsilon_0$ for $\xi = 0$ (left), $\omega - \epsilon_0$ marginalized on $\xi$ (middle) and $\xi - \epsilon_0$ with $\omega = -1$ (right). 
The are drawn for $\Delta \chi^2 = 2.30$ and $6.18$.}
\label{fig:qzw}
\end{figure*}

Figure \ref{fig1} shows the behaviour of the solutions by fixing the values
of two parameters and varying the third. The left panel corresponds
to $\omega=-1$, $\xi=0.2$ and different values of $\epsilon_{0}$. In this case, a mix of baryons 
($\lesssim 20\%$) and dark matter ($\gtrsim 80\%$) dominates the past evolution of the Universe whereas the 
dark energy is always the dominant component
from a value of $a \lesssim 1$ on. However, largest negative values of 
$\epsilon_{0}$ produce \textquotedblleft{}forbidden\textquotedblright{} solutions with negative density of dark energy. 
The middle panel corresponds to $\omega=-1$, $\epsilon_{0}=0.1$ and different $\xi$. Note that, although all the 
cosmological solutions are currently accelerated (as recently indicated by SNe Ia data), models with negative 
values of $\xi$ fail to reproduce the past dark matter-dominated epoch, whose existence is fundamental for the 
structure formation process to take place. In this case, the dark matter density vanishes at high-$z$ and the 
Universe is fully dominated by the baryons (for a CMB analysis in a baryon-dominated universe, [see (Griffiths, Melchiorri \& Silk 2001)].
The right panel corresponds to $\xi=0.2$, $\epsilon_{0}=0.1$ and different $\omega$, respectively. The behaviour 
of the system
is quite robust and the solutions for $N>0$ always converge to the fixed point $C_{1}$, regardless of the values 
of the model parameters. On the other hand, the state of the system at earlier times ($N<0$) is strongly dependent 
on the values of $\xi$ and, to a lesser extent, on the values of $\epsilon_{0}$. 

We have also verified that the introduction of small variations of the density conditions at $N=-6$ produces a shift 
of the time at which the dark energy density
starts to dominate over the other two components, but the behaviour of the solutions is qualitatively the same and the 
fixed point $C_{1}$ represents a global attractor of the system. 

In Fig. \ref{fig2}, we show the 3D ($X,Y,Z$) and 2D ($X,Y$) phase spaces for $\xi=0.2$, $\omega=-1$, 
$-0.5\leq\epsilon_{0}\leq0.5$ (left),
$\epsilon_{0}=0.1$, $\omega=-1$, $-0.5\leq\xi\leq0.5$ (middle),
and $\epsilon_{0}=0.1$, $\xi=0.2$, $-1.2\leq\omega\leq-0.8$ (right).
Note that all trajectories, even those starting close to the fixed
point $C_{2}$ converge to $C_{1}$. Thereby, the coupling between dark matter and dark energy should lead to 
an accelerated scaling attractor solution as required to solve the coincidence problem.

We may conclude that, within the 
considered range of variation of the model parameters, the Universe 
will always evolve to a phase currently dominated by dark energy.
Even if the fixed point $C_{1}$ becomes unstable at future times
($N>6$), the Universe will remain there forever unless a perturbation
is added to the model.

\section{Observational analysis}

Now, we will discuss the observational aspects of this class of interacting models. To this end, let us first write the Friedmann equation as
\begin{equation}\label{wang}
H^{2}= \frac{8\pi G}{3}(\rho_{b} + \rho_{dm} + \rho_{x})\;,
\end{equation}
where the evolution of the components $\rho_{b}$, $\rho_{dm}$ and $\rho_{x}$ can be found from Eqs. (\ref{def5}), (\ref{inicial}), (\ref{dmatter}) and (\ref{baryons}).

In order to delimit the bounds on $\omega$, $\epsilon_0$ and $\xi$ parameters, we use different observational sets of data as the most recent SNe Ia compilation, the so-called Union 2.1 sample compiled by (Suzuki et al. 2012) which includes 580 data points after selection cuts. The best fit of the parameters is found by using a $\chi^2$ statistics, i.e.,
\begin{equation}\label{chisquare}
\chi^2_{\rm{SN}} = \sum_{i=1}^N\frac{{[\mu_{p}^{i}(z|s)(z_i) - \mu_{o}^{i}(z|s)}]^{2}} {\sigma_i^2}\;,
\end{equation}
where $\mu_{p}^{i}(z|s) = 5\log d_L + 25$ is the predicted distance modulus for a supernova at $z$, $d_L$ is the luminosity distance, $\mu_{o}^{i}(z|s)$ is the extinction corrected distance modulus for a given SNe Ia at $z_i$ and $\sigma_i$ is the uncertainty in the individual distance moduli.

Additionally, we also use measurements derived from the product of the CMB acoustic scale 
\begin{equation}
f_{z_{BAO}} \equiv \frac{d_A (z_*)}{D_V(z_{\rm{BAO}})} \frac{r_s(z_d)}{r_s(z_*)}\;,
\end{equation}
where $d_A (z_*)$ is the comoving angular-diameter distance to recombination ($z_* = 1090$), $D_V$ is the dilation scale, $r_s$ is the comoving sound horizon at photon decoupling and $z_d \simeq 1020$ is the redshift of the drag epoch (at which the acoustic oscillations are frozen in). For $z_{\rm{BAO}} =$ 0.2, 0.35 and 0.6., one 
finds $f_{0.2} = 18.32 \pm 0.59$, $f_{0.35} = 10.55 \pm 0.35$ and $ f_{0.6} = 6.65 \pm 0.32$ (Sellerman et al. 2009; Blake et al. 2011) [see also (Percival et al. 2010)].

We perform a joint statistical analysis, by minimizing of the function $\chi^2_{\rm{T}} = \chi^2_{\rm{SN}} + \chi^2_{\rm{BAO/CMB}}$, where $\chi^2_{\rm{BAO/CMB}}$ correspond to the BAO/CMB $\chi^2$ function. In our statistical analysis we fix $\Omega_{b,0} = 0.0416$ and $\Omega_{dm} = 0.24$ which is in good agreement with current observational estimates and we also marginalize over the Hubble parameter $H_0$ the $\chi^2_{\rm{SN}}$ function.

\begin{table}[t]  
\begin{center}  
\caption{Limits on parameters and error bars at $1\sigma$.}
\begin{tabular}{rrrll}  
\hline  \hline \\
\multicolumn{1}{c}{Model}&
\multicolumn{1}{c}{\quad $\epsilon_{0}$}&
\multicolumn{1}{c}{ \quad $\omega $}& 
\multicolumn{1}{c}{$\xi$}\\ \hline \\
CQ $(\xi = 0)$ &-0.001 $\pm 0.07$& -1.01$ ^{+0.1}_{-0.1}$ \quad &  \quad \quad $-$\\ 
CQ  $(\xi \neq 0)$  & -0.005$^{+0.08}_{-0.08}$ & -1.05 $\pm 0.12$ & \quad \quad $-$\\
$\Lambda(t)$CDM & 0.002$^{+0.1}_{-0.1}$ &$-$ \quad \quad & 0.95$^{+0.18}_{-0.17}$\\ 
\hline  \hline
\end{tabular} 
\end{center} 
\end{table} 

Figure 3 shows the confidence regions (68.3\% and 95.4\% CL) in the planes $\omega - \epsilon_0$ for $\xi = 0$ (left), $\omega - \epsilon_0$ marginalized on $\xi$ (middle) and $\xi - \epsilon_0$ with $\omega = -1$ (right) obtained from the joint analysis described above. We see that in all panels both negative and positive values for the interacting parameter are allowed by these analyses. Physically, this amounts to saying that not only an energy flow from dark energy to dark matter ($\epsilon_{0} > 0$) is observationally allowed but also a flow from dark matter to dark energy ($\epsilon_{0} < 0$). In middle panel of Fig. 3 we see that both quintessence and phantom behaviours 
are acceptable regimes. In right panel of Fig. 3 we show the analysis for $\epsilon_{0}$ and $\xi$ with the dark energy EoS fixed at $\omega = -1$. As expected, we note that the current observational bounds on $\xi$ are quite weak since it appears as a power of the scale factor in the coupling function. However, it is interesting to observe that when $\xi$ takes more negative values $\epsilon_0 \rightarrow 0$ making the interaction between the dark components vanish. Table 1 shows a summary of the main results of our observational analyses.

\section{Conclusions}

We have investigated a general class of models with interaction between dark matter and dark energy 
in which the coupling in the dark sector is a power law of the cosmic scale factor ($\propto a^{\xi}$) 
and the EoS parameter may take any value $\omega < 0$.

We have also studied the dynamical behavior of this general class of models and we have shown from numerical 
simulations that, for a large set of parameter values ​​that characterize this class of models, the currently accelerated 
regime is preserved. This behaviour may alleviate the coincidence problem. 

We have also performed a joint statistical analysis using recent data of SNe Ia (Union 2.1) togther with 
the so-called BAO/CMB ratio at three redshifts, $z = 0.2$, $z = 0.35$ and $z = 0.6$ in order to constrain 
the free parameters of this class of interacting models. Best fits are obtained for a weak coupling 
($\propto 10^{-3}\rho_{dm}$) with values of $\omega \sim -1$ and $\xi \sim 0.9$.

\begin{acknowledgments}

F. E. M. Costa acknowledges financial support from CNPq (Brazilian Research Agency) grant no. 453848/2014-1 and UFERSA No 8P1618-22. A. O. Ribeiro 
and F. Roig acknowledge financial support from CAPES (Brazilian Graduate Studies Agency) and CNPq.

\end{acknowledgments}


\label{lastpage}
\end{document}